\documentclass{IOS-Book-Article}

\usepackage{mathptmx}
\usepackage{ulem}
\usepackage{soul}\setuldepth{article}
\usepackage{cite}
\usepackage{comment}
\usepackage{tabularx}
\usepackage{multirow}
\usepackage{booktabs}
\usepackage[table,xcdraw]{xcolor}
\usepackage{subcaption}

\usepackage{tikz}

\usepackage{enumitem}

\usepackage{todonotes}
\newcommand{\mac}[1]{{\color{red}#1}}

\def\hb{\hbox to 11.5 cm{}}

\begin{document}

\pagestyle{headings}
\def\thepage{}
\begin{frontmatter}              

\title{Neuro-Symbolic Query Optimization in Knowledge Graphs}

\markboth{}{July 2024\hb}

\author[A]{\fnms{Maribel} \snm{Acosta}\orcid{0000-0002-1209-2868}%
\thanks{Corresponding Author: Maribel Acosta, maribel.acosta@tum.de}},
\author[A,B]{\fnms{Chang} \snm{Qin}\orcid{0009-0002-4781-2391}}
and
\author[A]{\fnms{Tim} \snm{Schwabe}\orcid{0009-0009-7957-603X}}

\runningauthor{Acosta et al.}
\address[A]{Technical University of Munich}
\address[B]{Fraunhofer-Institut für Intelligente Analyse- und Informationssysteme IAIS}

\begin{abstract}
This chapter delves into the emerging field of neuro-symbolic query optimization for knowledge graphs (KGs), presenting a comprehensive exploration of how neural and symbolic techniques can be integrated to enhance query processing. Traditional query optimizers in knowledge graphs rely heavily on symbolic methods, utilizing dataset summaries, statistics, and cost models to select efficient execution plans. However, these approaches often suffer from misestimations and inaccuracies, particularly when dealing with complex queries or large-scale datasets. Recent advancements have introduced neural models, which capture non-linear aspects of query optimization, offering promising alternatives to purely symbolic methods.
In this chapter, we introduce neuro-symbolic query optimizers, a novel approach that combines the strengths of symbolic reasoning with the adaptability of neural computation. We discuss the architecture of these hybrid systems, highlighting the interplay between neural and symbolic components to improve the optimizer's ability to navigate the search space and produce efficient execution plans. Additionally, the chapter reviews existing neural components tailored for optimizing queries over knowledge graphs and examines the limitations and challenges in deploying neuro-symbolic query optimizers in real-world environments. 
\end{abstract}

\end{frontmatter}
\markboth{July 2024\hb}{July 2024\hb}

\section{Introduction}
\label{sec:introduction}
Knowledge graphs (KGs) and other data models can be interrogated using queries formulated in a query language. 
For example, for knowledge graphs modeled with the Resource Description Framework (RDF), queries are formulated using the SPARQL query language. 
To compute the answer for a given query, a query processor comprises different components that implement strategies to speed up the query execution while producing correct results. 
Such strategies include resorting to access methods to retrieve portions of the stored dataset efficiently, re-ordering the operators that occur in the query to minimize the number of intermediate results, or implementing specific algorithms to implement operators that produce intermediate results quickly.   

In query processing, the optimizer is the component that selects the strategies to execute a given query over a dataset. 
These strategies are encoded in a query plan, which is represented as a tree. 
The leaves of this tree correspond to access methods to retrieve relevant data from the dataset, while the internal nodes are operators to combine intermediate results and produce the final query answer.  
Formally, the query optimization problem can be defined as a search problem~\cite{chaudhuri1998overview} whose goal is to identify a plan from the space of possible plans that minimizes some objective function, e.g., the query runtime or latency.  
Ibaraki and Kameda~\cite{ibaraki1984optimal} formally demonstrated that identifying the optimal plan is an NP-complete problem. 
To explore the search space and identify efficient plans, traditional query optimizers implement symbolic components based on dataset summaries, statistics, cost models, (algebraic) rules, or heuristics. 

However, even with detailed statistics and advanced techniques, symbolic query optimizers are prone to errors due to misestimations or inaccurate models. 
Producing sub-optimal plans has tremendous effects on query latency, where execution times can be affected by orders of magnitude or even prevent a query from being answered in a reasonable amount of time. 
This problem is exacerbated when evaluating complex queries or dealing with very large KG. 
To overcome these limitations, neural models for query optimization have been proposed for query engines over relational databases and knowledge graphs. 
Most of these solutions focus on replacing a specific component in the query optimizer by using machine learning models based on neural networks or deep learning, as they are able to capture complex relationships of the factors that affect the query performance, e.g., data correlations, complexity of the query operators, and (hardware) resource usage. 
This chapter goes a step beyond the literature and introduces the notion of neuro-symbolic query optimizers, where we discuss the integration of neural and symbolic components during query optimization. 
Furthermore, this chapter provides details of existing solutions for neural components to optimize queries over knowledge graphs. 
We finally present a discussion of existing limitations and future directions of neuro-symbolic query optimizers.

The remainder of this chapter is structured as follows. 
\S\ref{sec:preliminaries} introduces the preliminaries, including symbolic methods for query optimization and the general architecture of query optimizers. 
\S\ref{sec:databases} briefly presents existing neural and neuro-symbolic methods for relational databases. 
\S\ref{sec:ns-optimizer-kg} describes the architecture of neuro-symbolic optimizers for knowledge graphs. 
\S\ref{sec:future_directions} discusses challenges and future directions for deploying neuro-symbolic optimizers in real-world systems. 
\S\ref{sec:conclusion} concludes this chapter.

\section{Preliminaries}
\label{sec:preliminaries}

\subsection{Symbolic Methods for Query Optimization}

Symbolic query optimization typically refers to a type of optimization that involves manipulating and reasoning about the symbolic representation of queries and their plans. 
This approach focuses on theoretical and logical transformations. 
The database literature distinguishes three types of optimizers related to symbolic query optimization: heuristics-based, rule-based, and cost-based optimization.

\paragraph{Heuristic-based Optimization}
These optimizers use heuristics and `rules of thumb' to make decisions about query plans. They rely on strategies or heuristics to simplify and speed up the optimization process. For example, a heuristic might dictate that certain subplans should always be preferred. These optimizers are generally fast (i.e., polynomial time), but they do not always find the optimal plan.

\paragraph{Rule-based Optimization}
These optimizers use predefined rules to devise plans. In contrast to heuristics, these rules are typically based on the properties of the query workload and the schema of the database at hand. For instance, a rule might specify that using a specific access method is preferable for certain query types. Rule-based optimizers are straightforward to implement, are predictable, and can generate plans fast. 
Yet, they suffer from similar limitations as heuristics-based optimizers. 

\paragraph{Cost-based Optimization} 
These optimizers explore a larger space of possible plans for a query while estimating the cost of each plan. 
The optimizer then selects the plan with the lowest estimated cost. 
Cost-based optimizers are generally more accurate in finding the optimal plan because they assess more plans and consider their execution costs, though they can be more computationally expensive than other optimizers.

\smallbreak
From these methods, rule-based optimizers are most directly related to symbolic query optimization due to their reliance on symbolic transformation rules. 
Heuristics-based optimizers also use symbolic reasoning but in a more practical way. 
Lastly, cost-based optimizers focus more on empirical cost analysis, yet they still perform some symbolic manipulations by exploiting algebraic transformations over query expressions to preserve the equivalence of plans.

\begin{figure}[t!]
    \centering
    \includegraphics[width=\textwidth]{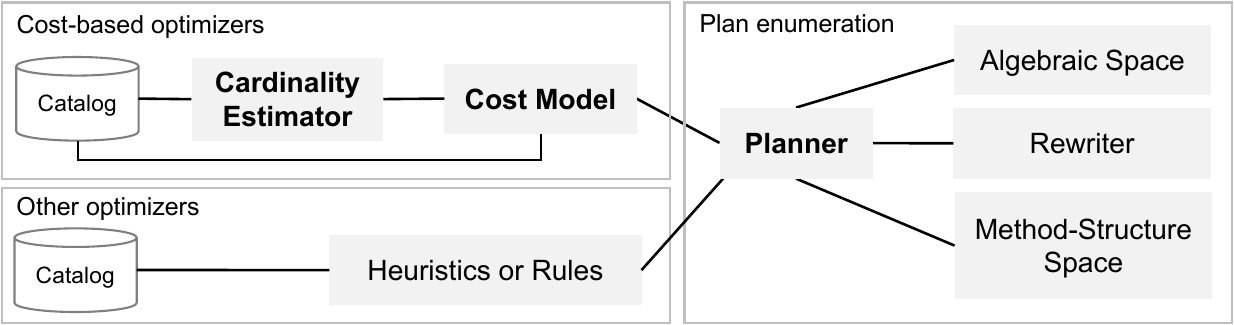}
    \caption{Simplified architecture of query optimizers}
    \label{fig:architecture}
\end{figure}
\subsection{General Architecture of Query Optimizers}
Figure~\ref{fig:architecture} outlines the components of query optimizers. 
The cost-based optimizer components include the cardinality estimator~(\S\ref{sec:cardinality_estimator}) and the cost model~(\S\ref{sec:cost_model}). 
The planner (\S\ref{sec:planner}) explores potential plans to identify the optimal execution strategy using the cost model (in the case of cost-based optimizers) or pre-defined strategies encoded as heuristics or rules (in the case of other symbolic optimization techniques).

\subsubsection{Cardinality Estimator}
\label{sec:cardinality_estimator}
This component determines the cardinality or number of results produced by a subplan. 
In query optimization, the cardinality of a subplan is typically assumed to be directly related to the cost of executing such subplan. 
Essentially, the more results there are to process, the more expensive the execution of the subplan is in terms of time and space. 

To perform cardinality estimation, the optimizer relies on statistical summaries of the dataset stored in the  catalog.  
These summaries typically include relation cardinalities, data distributions, and other relevant statistics captured in histograms, data samples, and even pre-computed results for frequent query patterns. 
However, even with detailed statistics, cost-based optimizers can produce inaccurate  estimations. 
This is because they often assume that attribute values are independent and uniformly distributed, which allows for simplifying the process of computing the statistics, performing fast cardinality estimations, updating the summaries, and generalizing the optimization techniques to several datasets. 
However, these assumptions do not always hold in practice~\cite{deshpande2007adaptive}, thus, requiring advanced techniques such as multivariate statistics, enhanced histograms, and even machine learning. 
The latter will be the focus of this chapter. 

In some scenarios, accurate statistical summaries of the dataset are not available. For example, in query processing over remote data sources, it may not be possible to gather sufficient statistics. Another example is in query processing over dynamic data, where the dataset changes frequently, making it nearly impossible for the optimizer to maintain up-to-date statistics. For cases like this, the optimization process needs to be complemented with adaptive query processing techniques~\cite{deshpande2007adaptive,acosta2011anapsid,acosta2015networks} to adjust the plan execution based on the performance observed during runtime.


\subsubsection{Cost Model}
\label{sec:cost_model}
In cost-based optimization, the cost model is used to estimate the resources required to execute query plans. 
In this context, the cost is a numerical value which is associated with each plan. 
Although this cost does not necessarily represent the exact runtime of a query, it is correlated with it: a higher cost indicates a longer query runtime.

Different database systems implement different models or formulas to estimate the cost of a plan. 
Still, a typical cost model formula considers several factors, such as I/O cost, CPU cost, and sometimes memory usage. 
For example, for a plan that involves scanning a table and applying a filter, the cost might be estimated as the linear combination of number of sequential pages read from disk plus the number of tuples processed. 
The following is an example of such a cost model implemented in PostgreSQL~\cite{postgresql_explain}:  
\begin{equation}
\textit{Cost}(P) = 
\overbrace{(C_{seq\_page} \cdot \underbrace{\#disk\_pages\_read)}_{\textit{Catalog}}}^{\textbf{I/O Cost}} + 
\overbrace{(C_{cpu\_tuple} \cdot \underbrace{\#tuples\_scanned}_{\textit{Cardinality estimator}})}^{\textbf{CPU Cost}}
\label{eq:costmodel}
\end{equation}

As shown in Figure~\ref{fig:architecture}  and Eq.~(\ref{eq:costmodel}), the cost model relies on the dataset statistics stored in the catalog and the cardinality estimator to compute the efficiency of a plan.  

In practice, developing a cost model that accurately reflects the true execution cost of different query plans can be complex. 
Finding suitable values for the constants $C$ used in a cost model is not straightforward as these are coupled to the database configuration, the query workload diversity, and even the hardware where the database is deployed. 
Furthermore, all of these factors can change over time due to changes in the dataset, queries, and technical infrastructure.  
Therefore, finding optimal values for constants and other parameters in a cost model requires a combination of theoretical knowledge, empirical analysis, and continuous tuning to ensure the cost model remains accurate over time. 

\subsubsection{Planner}
\label{sec:planner}
For a given query, the planner traverses the search space of possible plans by comparing \textit{equivalent plans}, i.e., plans whose executions produce the same results. 
The space of equivalent plans is given by~\cite{ioannidis1996query}: 
(i)~the \textit{algebraic space}, which is the set of rules that preserve plan equivalence [36] (e.g, operator commutativity and associativity), 
(ii)~the \textit{rewriter} applies simple transformations to the query to produce more efficient plans, e.g., by using views, flattening out subqueries, redundant subexpression elimination, etc.   
and (iii)~the \textit{method-structure space} that contains the available implementations of access methods and logical operations specified in the query.  

During the plan traversal, the planner compares equivalent plans to find an optimal plan. 
For this, planners may implement different search strategies [86, 96], including exhaustive search using Dynamic Programming [144], greedy algorithms that prune large sub-spaces of plans and typically run in polynomial time, randomized algorithms that generate an initial plan randomly which is later refined, and hybrid search strategies such as Iterative Dynamic Programming (IDP) [96] combining DP with greedy steps.  

In cost-based optimization, the planner relies on the cost model to guide the exploration of plans and identify the best ones. 
Similarly, optimizers with different architectures may also implement search strategies like the ones discussed here, but they use heuristics or rules to enumerate and compare equivalent plans.

\section{Literature in Neural Components in Relational Databases}
\label{sec:databases}


In the context of neural approaches for relational databases, the cardinality estimators, cost models, and plan traversal components are enhanced with learned models. 
Each of these components benefits from learning different aspects of query processing via machine learning and neural networks. 
In the following sections, we will explore how the learned models are integrated into optimizers to improve the query performance. 

\subsection{Learned Cardinality Estimation}
\label{sec:learned_cardinality_estimation}
Cardinality estimators that rely on machine learning may follow two paradigms: the query modeling (\S\ref{sec:query_model}) and the data modeling (\S\ref{sec:data_model}). 
Query modeling approaches learn a mapping function between a query or a plan and its cardinality, treating cardinality estimation as a regression problem. 
Instead, data modeling treats cardinality estimation as a density estimation problem by focusing on learning joint data distributions~\cite{Sun2021Learned}. 
An overview of existing solutions for learned cardinality estimations is shown in Table~\ref{tab:card_est_db}.

\begin{table}[t!]
\caption{Overview of learned cardinality estimation techniques for databases. Adapted from \cite{zhu2024learned}}
\begin{tabularx}{\textwidth}{p{2cm} X p{5.9cm}}
\toprule
\textbf{Category}                       & \textbf{Learning Method}                                      & \textbf{Techniques}                                                                                                                                                                                                           \\ \midrule
                               & \cellcolor[HTML]{EFEFEF}Statistical Model            & \cellcolor[HTML]{EFEFEF}Linear Model; Tree-based Ensemble; XGBoost; Mixture Model                                                                                                                                    \\ 
\multirow{-4}{*}{\textbf{Query-Driven}} & NN-Based                                            & Fully Connected Neural Network; Multi-set Convolutional Network; Adding Pooling Layers; Adding Containment Rate; Query Masking; Segmentation; Ensemble of Deep Models; Bayesian Deep Learning; Query Re-optimization \\ \midrule
                               & \cellcolor[HTML]{EFEFEF}Kernel-Based                 & \cellcolor[HTML]{EFEFEF}Kernel Density Function                                                                                                                                                                      \\ 
                               & Auto-Regressive Model                                & Adding Gaussian Mixture Models; Single Table; Multi-Table                                                                                                                                                            \\ 
                               & \cellcolor[HTML]{EFEFEF}Probabilistic Graphic Model & \cellcolor[HTML]{EFEFEF}Bayesian Network; Revitalized Bayesian Network; Sum-Product Network; FSPN, Factor Graph and Join Histogram                                                                                   \\ 
\multirow{-6}{*}{\textbf{Data-Driven}}  & Others                                               & Normalizing Flow; Summarization Models                                                                                                                                                                               \\ 
\midrule 
\textbf{Hybrid}                         & \cellcolor[HTML]{EFEFEF}Others                       & \cellcolor[HTML]{EFEFEF}Deep Auto-Regressive Models; Attention on Transformer Models                                                                                                                                 \\ \bottomrule
\end{tabularx}
\label{tab:card_est_db}
\end{table}

\subsubsection{Query-Driven Models} 
\label{sec:query_model}
Query-driven models are applied when the full dataset is unavailable or too large to learn from. 
These approaches use query logs to predict the cardinality of the new queries with similar patterns and, therefore, it is regarded as a supervised learning process. 


Query-driven modeling approaches perform (1) a 
\textit{sampling phase} to extract information related to schema and attribute values tables, columns from the schema, values from each column for predicates~\cite{Sun2021Learned}, or the physical plans generated by the optimizer \cite{sun2019end} are gathered. 
To pass this information to a neural network, the approaches build a (2) \textit{query featurization} where the operations included in the query are transformed into numerical vectors, e.g., binary or one-hot encoding vectors. 
Based on the query features, the approaches learn (3) a \textit{regression model} to predict the cardinality. 
As for model selection, neural approaches are preferred because the data can have non-linear, complex distributions in general \cite{dutt2019selectivity}.
Many customized models based on neural networks have been investigated and proven to outperform traditional cardinality estimators. 
These models include Multi-set Convolutional Networks (MSCNs)~\cite{kipf2018learned}, Recurrent Neural Networks (RNNs)~\cite{ortiz2019empirical}, and Fully Connected Networks (FCNs)~\cite{kim2022learned}. 

Query-driven methods are proven to accurately predicate cardinalities, but only when the distributions of training and test queries are similar~\cite{kim2022learned}. Besides, collecting queries for the training data and executing them on real databases is also expensive. Given these limitations, data-driven approaches for learning cardinality estimation are widely investigated. 


\subsubsection{Data-Driven Models} 
\label{sec:data_model}
These models learn to predict the query cardinality from data distributions by learning a joint data distribution of each data point in the dataset~\cite{Sun2021Learned}. 
Existing modeling approaches greatly differ in the model selection, as shown in Table~\ref{tab:card_est_db}. 

In the following, we briefly discuss unsupervised methods for data-driven modeling that are neural-based. 
Prominent approaches based on neural networks are Sum-Product Networks (SPNs) and autoregressive models. 
SPNs~\cite{hilprecht2019deepdb} split data into clusters and groups based on similarity and correlation, respectively, using SUM and PRODUCT operators to efficiently compute joint data distributions and cardinalities. 
Autoregressive models treat each tuple or row as a sequence \cite{yang2019deep, hasan2020deep}. To predict the value of the next attribute in the tuple, it factorizes the joint distribution and computes conditional distributions based on the preceding attributes. 
It then aggregates these probabilities for data samples matching a query to estimate the query's cardinality~\cite{Sun2021Learned}.

Although these approaches can achieve high accuracy, their inference time is much longer because of the adoption of more specialized models. 
For example, the total time of Naru~\cite{yang2019deep} is sensitive to the running device, which needs 5ms to 15ms on GPU, whereas CPU can be up to $20\times$ slower, which will be a blocking issue of bringing these accurate data-driven unsupervised learned estimators into production~\cite{wang2020we}.

\subsection{Learned Cost Models} 
Learned cost models based on neural~(\S\ref{sec:neural_cost_model}) and neuro-symbolic~(\S\ref{sec:ns_cost_model}) approaches have been widely investigated and proven to outperform traditional methods \cite{boulos1997neural,Sun2021Learned,marcus2019neo,lan2021survey,marcus2019plan,yang2023rethinking}. 

\subsubsection{Neural Cost Models}
\label{sec:neural_cost_model}

Neural–based approaches to learned cost models are mainly supervised methods and  are applied to replace cost models by predicting performance metrics. The input is generally presentations of queries, operators or other required characteristics which are fed into the neural networks. NN-based \cite{boulos1997neural}, tree convolution\cite{marcus2019neo, marcus2019plan} and tree-structured deep neural network \cite{sun2019end} are used to capture the complex relationships and patterns in query plans. As for output,  various performance metrics, such as latency, are predicted \cite{marcus2019neo, marcus2019plan}.

\subsubsection{Neuro-symbolic Cost Models}
\label{sec:ns_cost_model}
Although learning-based approaches with neural networks for cost estimation have been intensively studied these years, the inherent limitations, such as training overhead, poor generalization among databases as well as the lack of explainability for end-to-end predictions, are unavoidable~\cite{yang2023rethinking}. Therefore, traditional formula-based cost models --as the example from Eq.~(\ref{eq:costmodel})-- have now improved with advanced learning-based models \cite{yang2023rethinking}. 

Instead of manually setting the values for constants and parameters in formula-based cost models, the neuro-symbolic cost models learn optimal settings offline.  
As new data and query workloads come into the system, parameters are further refined and adjusted dynamically with online learning to adapt to changing conditions or configurations. 
In this way, transferability can also be achieved by a lightweight learning scheme. Besides, inherent advantages of formula-based cost models are utilized, such as training is based on the existing knowledge of the cost model and the interpretability of the cost model, which makes the training process more efficient and interpretable~\cite{yang2023rethinking}. 

\subsection{Learned Plan Traversal}
\label{sec:learned_plan_traversal}
One of the main challenges in cost-based query optimization is that the plan enumeration phase is sensitive to errors in the cardinality estimations and the cost model. 
To avoid this, learned plan traversal approaches skip these aspects, and learn a model to effciently identify the subspace of plans where optimal plans may reside.

Current learned planners are neural-based using deep reinforcement learning (DRL)~\cite{lan2021survey}.
With DRL, an agent learns to chose the policy which make the cumulative rewards maximun. 
In the context of query optimization, the state and action of reinforcement learning are the current subplans and the combination of two subplans into a new plan. 
In this way, the planner learns a decision policy to map states to actions, with the maximum expected reward~\cite{krishnan2018learning}. Depending on learning from past queries or during the execution of the current query, learned plan travesal can be categorized into offline-learning and online-learning methods.

\paragraph{Offline Learning} 
Methods like Q-learning~\cite{krishnan2018learning, ramadan2022rl_qoptimizer}, Deep Q-Network \cite{heitz2019join}, and policy gradient methods\cite{marcus2018deep} are implemented to develop policies based on prior experiences, which include actions, states, and rewards derived from previous queries. The query plans (trees), as states, can include information about tables and operation selections~\cite{krishnan2018learning, marcus2018deep} as well as the structure plan's with tree-structured long short-term memory~\cite{rtos}. 

While offline learning can predict better operation orders, they also exist inherent limitations. 
The main shortcoming of these methods is that new queries must be similar to prior queries to make the prediction effective, and this similarity must be recognizable~\cite{trummer2021skinnerdb}. 
In contrast, online learning offer adaptive strategies that continuously update their models based on new queries, as explained in the following. 

\paragraph{Online Learning} 
With online-learning methods, the query optimizer learns to find optimal operator orders at runtime. 
This approach was first investigated through an adaptive query processing mechanism known as Eddies. 
While Eddies are not neural-base, it inspires the subsequent work to improve adaptive query processing using Reinforcement Learning~\cite{tzoumas2008reinforcement}. 
The fundamental idea treats the query execution process as a training episode, whereas a more advanced approach~\cite{trummer2021skinnerdb} splits the execution into many small time slices, during each of which it executes a selected subplan, trains the Upper Confidence Bounds to Trees based on the subplan's real-time performance, and finally, merges the intermediate results produced in each time slice to generate the final result~\cite{zhou2020database}.

Fully online-learning and adaptive approaches introduces overhead, which may increase the runtime of individual queries~\cite{krishnan2018learning}. Besides, the complexity of fully online-learning query optimizer makes its integration into production systems difficult\cite{trummer2021skinnerdb}.

\section{Towards Neuro-Symbolic Optimizers for Knowledge Graphs}
\label{sec:ns-optimizer-kg}
The problem of query optimization and solutions for knowledge graphs (KGs) takes a lot of inspiration from the ones developed for relational databases. 
However, existing data models for KGs present fundamental properties that greatly differ from the relational model. 
This hinders the direct application of database techniques to KGs. 
In the context of query optimization, three of these \textbf{fundamental properties of KGs} are:  

\begin{enumerate}
    \item Connectedness: KGs are graph-based structures where entities are represented as nodes, which are interconnected via edges that represent relationships between them.  In graph models such as RDF, where the atomic data structure is a triple, the relationships or connections in the KG are considered first-class citizens.  

    \item Semi-structuredness: KGs follow a schema-less paradigm without imposing the same structures or restrictions over the entire datasets. Due to their schema-less nature, KGs often present irregular data structures where some nodes are over-specified with multiple attributes or predicates while others are highly incomplete. 

    \item Unexpected data correlations: The data distributions in real-world KGs follow a power law, e.g., a small number of nodes have a high degree of connections, while most nodes have a few connections. 
    This is very common in RDF graphs~\cite{DBLP:journals/semweb/ZlochAHCD21} due to the presence of frequent properties defined in RDF/S and OWL.  

\end{enumerate}
The properties (1), (2), and (3) make the process of learning patterns and joint data distributions from the KG very challenging. 
In particular, property (3) impedes applying techniques from relational databases, where data is typically more uniformly distributed.

Due to these fundamental differences, neural and neuro-symbolic methods for query optimization have been proposed for KGs. 
In this chapter, first, we present a characterization of neuro-symbolic architectures for query optimizers (\S\ref{sec:ns_query_optimizer}). 
Then, we describe how the components of a KG query optimizer can implement  
learned cardinality estimation~(\S\ref{sec:kg_learned_cardinality_estimation}), 
learned cost models~(\S\ref{sec:kg_learned_cost_models}), 
and learned planned traversal~(\S\ref{sec:kg_learned_plan_traversal}).

\begin{figure}[t!]
    \centering
    \includegraphics[width=1\textwidth]{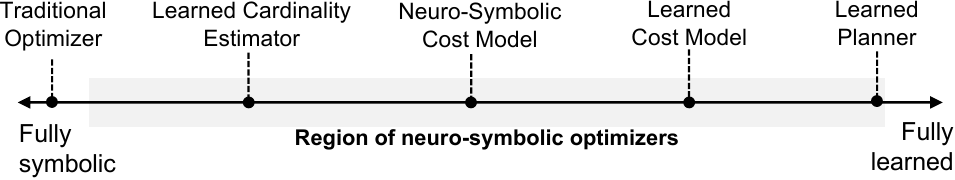}
    \caption{Spectrum of integrating learned components into a query optimizer}
    \label{fig:kg_spectrum}
\end{figure}

\subsection{Architecture of a Neuro-Symbolic Query Optimizer}
\label{sec:ns_query_optimizer}

Query optimizers can integrate learned or neural components in different ways. 
To characterize this, we present in Figure~\ref{fig:kg_spectrum} a spectrum comprising  a progression from traditional, fully symbolic query optimizers to fully learned, neural network-based optimizers, with various hybrid neuro-symbolic approaches in between.

\paragraph{Fully Symbolic or Traditional Optimizer} 
This represents the classic, rule-based query optimizer found in most systems (cf.  \S\ref{sec:preliminaries}). 
It relies entirely on symbolic methods, such as dataset statistics, predefined rules, and heuristics to make optimization decisions.

\paragraph{Fully Learned Optimizer} At the far end of the spectrum, this represents a fully learned query optimizer that relies entirely on neural networks or other machine learning models to perform all aspects of query optimization, from cost estimation to plan selection.

\paragraph{Region of Neuro-Symbolic Optimizers}
This middle area represents a blend of symbolic and neural approaches. It includes optimizers that combine learned models with traditional symbolic components, including: 

\begin{itemize}
    \item Learned Cardinality Estimator: A neural model is used to estimate the cardinalities of intermediate query results. 
    \item Neuro-Symbolic Cost Model: Integrates a learned model into the cost estimation process, either by learning the weights of the factors in the cost model or by combining neural estimates with traditional cost models. Currently, there are no approaches for neuro-symbolic cost models over KGs.  
    \item Learned Cost Model: The entire cost model is replaced by a learned model, which predicts the cost of executing query plans using machine learning techniques. 
    \item Learned Planner: The planning process itself is guided by a learned model, which suggests query plans based on data obtained from estimates or performance measurements obtained online or from past executions. 
\end{itemize}

The following sections focus on the architecture of components to build neuro-symbolic optimizers and discuss limitations and open challenges of existing solutions. 



\begin{figure}
    \centering
    \begin{subfigure}[t]{0.48\textwidth}
        \centering
        \includegraphics[width=\textwidth]{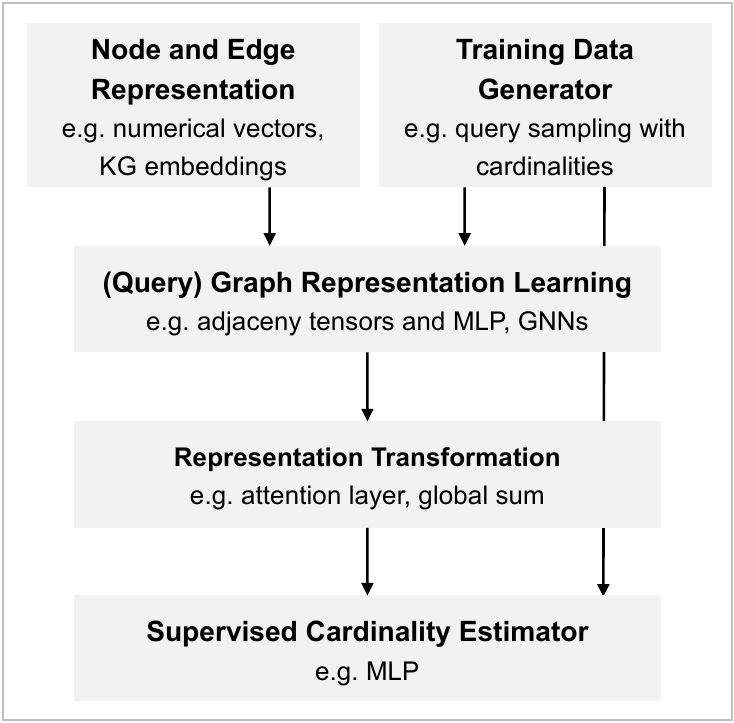}
        \caption{Query-driven models}
        \label{fig:kg_query_model}
    \end{subfigure}
    \begin{subfigure}[t]{0.48\textwidth}
        \centering
        \includegraphics[width=\textwidth]{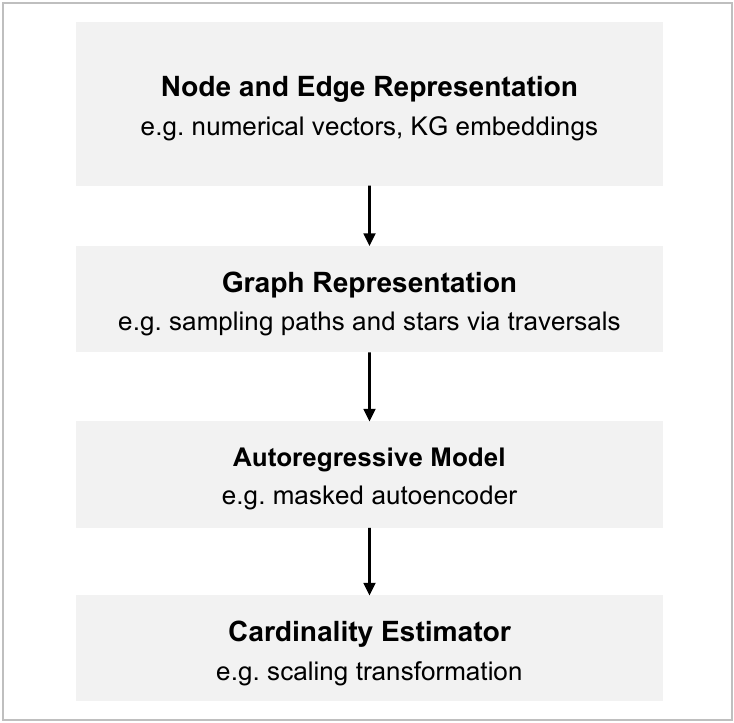}
        \caption{Data-driven models}
        \label{fig:kg_data_model}
    \end{subfigure}
    \caption{Neural architectures of learned cardinality estimators for knowledge graphs}
    \label{fig:kg_cardest_architecture}
\end{figure}

\subsection{Learned Cardinality Estimation}
\label{sec:kg_learned_cardinality_estimation}
Different architectures can be applied to learn cardinality estimations over KGs as shown in Figure~\ref{fig:kg_cardest_architecture}. 
In query-driven architectures (cf. Figure~\ref{fig:kg_query_model}), the model is trained in a supervised way using labeled training data, i.e., queries and their corresponding cardinalities. 
For this, the approaches first obtain a representation of the terms (nodes and edges) that occur in the queries and learn a representation for the queries themselves. 
These representations can be transformed, e.g., into a vector, and then passed to a supervised model to predict query cardinalities.  
In data-driven architectures (cf. Figure~\ref{fig:kg_data_model}), the model can be trained in an unsupervised way by learning the latent probability distributions of the nodes and edges in the KG. 
For this, the approaches traverse the KG to perform subgraph sampling. 
The sampled subgraphs are then used in an autoregressive model to learn the probabilities of the KG terms (nodes and edges); these probabilities are finally transformed into cardinality values by performing some scaling transformation.

\subsubsection{Node and Edges Representation}
\label{sec:vectors}
The first step is to represent the entities (nodes) and predicates (edges) that occur in the input data (i.e., queries or the KG itself) in a suitable way for neural models. 
This is done by transforming the nodes and edges into vector-based representations. 

\paragraph{Numerical Vector Encoding}  
A simple approach to represent entities and predicates is to assign them a numerical \textsc{id}, and then represent these \textsc{id}s with binary or one-hot encoding vectors. 
This representation is simple but does not include any information about the meaning or interconnections of entities or predicates in the KG, forcing the subsequent model to learn the semantics of these. 

\paragraph{Embedding-enhanced Representation}
To represent the individual entities and predicates that occur in the query, it is possible to use Knowledge Graph Embeddings (KGE), instead of assigning them arbitrary numerical \textsc{id}s. 
KGE can be trained from parts of the entire KG, and capture rich information about the entities and predicates, which can thus be predictive for cardinality estimation.  
This alleviates the need to encode this information into the subsequent model. 
Furthermore, using KGE has the advantage that some embedding algorithms are inductive, i.e., embeddings for new entities can be added incrementally without requiring retraining of the existing ones.

\subsubsection{Graph Representation and Cardinality Estimation}
\label{sec:graph}
Performing cardinality estimation over KGs is inherently different from the relational case. 
In KGs, both the dataset and the queries are graphs\footnote{Basic Graph Patterns in SPARQL correspond to subgraph patterns.}, therefore, the models greatly benefit  from representations that explicitly encode the connectivity of nodes and edges in the dataset.    
Multiple ways exist of encoding KGs and queries to suitably represent them to neural models to perform cardinality estimation, differing mainly in their expressivity.  



\paragraph{Adjacency Tensors with Multilayer Perceptrons}
This is a rather simple encoding to represent the input data using an adjacency matrix.
This representation can only be used in query modeling approaches, as the adjacency matrix of the entire KG is typically a very large and sparse structure. 
To represent a query, it is sufficient to use three matrices~\cite{LMKG}.
One is an adjacency tensor, to represent the graph structure of the query. 
The size of this tensor is fixed a priori with the maximum number of edges allowed by the model in any query. 
The other two matrices are also of fixed size to encode the entities and predicates that occur in a query, respectively, using vectors or embeddings as detailed in \S\ref{sec:vectors}. 
These three matrices can be flattened into vectors and processed using Multilayer-Perceptrons (MLP), combined, and transformed into a cardinality estimate using another MLP~\cite{LMKG}. 
This is a simple and efficient architecture, yet it has several drawbacks. 
First, the fixed dimensions of the MLPs implicate only queries with a (maximum) fixed number of nodes and edges can be processed by a given model, requiring training multiple models. 
Second, processing the adjacency tensor as a flattened vector using an MLP does not account for the permutation invariance of (query) graphs. 
That is, a permutation of the entities in the adjacency will still represent the same query graph but will look vastly different to the model. 
This requires the model to have more training data to learn this invariance, or to compute query canonical representations where permutations of the same query always produce the same representation.

\paragraph{Graph Neural Networks}
Instead of representing the KG or the graph queries using matrices, a fitting approach is to compute a representation using a Graph Neural Network (GNN). 
While this approach can also be applied to the entire KG, it can be challenging to scale up to very large KGs. 
Therefore, existing approaches instead apply it to query graphs~\cite{GNCE,LSS}.   
Using GNNs to represent graph structures and their connectedness has two main advantages. 
First, a single GNN model can process arbitrary query shapes and sizes by default, so there is no need for multiple models.
Second, GNNs are permutation-invariant and thus capture commutative of joins or \textsc{and} operations in query graphs, which makes these models more data efficient. 
The learned node representation with the GNN can then be combined, e.g., with a global sum to perform cardinality estimations using directly an MLP~\cite{GNCE} or an attention layer followed by an MLP~\cite{LSS}.

\paragraph{Autoregressive Models}
In the context of cardinality estimation, autoregressive models learn the correlation between the nodes and edges in the KG in an unsupervised way.
For this, the models perform random walks over the KG while traversing sub-structures with different shapes, e.g., stars that comprise several edges with the same subject or paths that are sequences of edges. 
One option is to use a masked autoencoder that predicts the probability of each term in the representation~\cite{LMKG}. 
This architecture is efficient since, by masking, all probabilities can be calculated in a single model, instead of relying on sequential processing. 
The estimated probability of the represented data can then be turned into a cardinality, e.g., by multiplying it by the size of the KG. 
Nonetheless, the main limitation of this approach is that this architecture is tightly coupled to the sampled structures. 
In practice, autoregressive models can capture paths or stars for cardinality estimation, yet it is difficult to implement them for more complex subgraphs~\cite{LMKG}.

\subsubsection{Limitations and Open Challenges}

Current neural models for cardinality estimation show impressive results in terms of accuracy, but they still present several limitations. 
First, while the studied models are rather small in terms of parameters, training on large corpora still requires extensive time and computational resources. 
In particular, supervised approaches require a representative and diverse query workload to be able to generalize to new queries. 
Still, real-world query workloads are not always accessible, and generating diverse queries is not necessarily straightforward and can also be challenging in terms of time and resource consumption. 
Second, while many methods show degraded yet acceptable performance in dynamic KGs, i.e.,  new  entities are added to the KG, substantial changes to the KG -- i.e., large amounts of added entities, schema changes -- require retraining the models, which hinders scalability. 
Hence, future work needs to develop more efficient models for rapid adaptability to dynamic data changes, e.g., embeddings or representations that are easily updatable and tailored to cardinality estimation as a downstream task.

\begin{figure}[t!]
    \centering
    \includegraphics[width=1\textwidth]{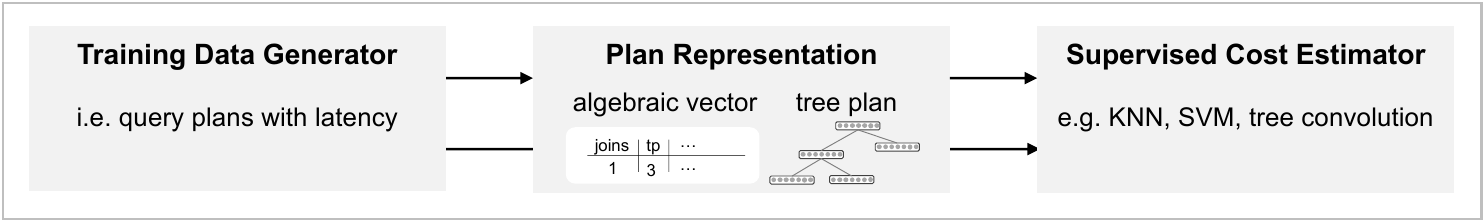}
    \caption{Architecture of learned cost models for KGs}
    \label{fig:kg_costmodel_architecture}
\end{figure}

\subsection{Learned Cost Models} 
\label{sec:kg_learned_cost_models}

Current learned cost models for KGs aim to directly predict the latency or runtime of a plan before execution. 
These models can be directly used by the query optimizer to aid the search for plans with minimal query latency. 
Compared to the relational case, models for KGs face the challenges of a possibly large number of operations in the query, resulting in a larger plan to be processed and difficulty in gathering concise statistics due to the semi-structuredness and unexpected data correlations that occur in KGs.
Figure~\ref{fig:kg_costmodel_architecture} shows the general architecture of learned cost models for KGs.

\subsubsection{Query Plan Representation}
In contrast to queries, query plan representations include the sequence of operations and other technical details about the execution strategies that the system will use to evaluate a query. 
To enable a trained model to predict the latency of a specific query plan, it is necessary to encode the plan into a format that the model can process, as discussed below.

\paragraph{Algebraic and Graph Similarity Vectors}
Query plans can be represented using fixed vectors that encode their characteristics. 
Algebraic features, including the number of triple patterns, operators and their order occurring in the plan, and depth of the query tree, can be used to coarsely characterize the plan~\cite{sparql_clustering_costmodel}. The structure of a query graph can be coarsely encoded by calculating its graph edit distances to a set of representative queries and using those distances as features~\cite{sparql_clustering_costmodel}.

\paragraph{Query Tree Representation}
Apart from these coarse representations, a plan can also be represented directly as a tree. 
The structure then explicitly encodes the operators and their order, and the nodes can be encoded as vectors that hold information such as the cardinality of the subtree, as well as nodes and edges instantiated in the query, etc.

\subsubsection{Cost Model Predictors}
Various machine learning architectures leverage the representation of query plans to estimate their latency accurately. 
Some architectures include simple models like k-nearest neighbors and support vector machines. 
More advanced methods, such as tree convolution, utilize the structural information inherent in query trees. These methods demonstrate the evolution from coarse representations to more sophisticated techniques that better capture the complexities of query execution plans.

\paragraph{K-Nearest Neighbors and Support Vector Machines}
Based on the algebraic and graph similarity vectors, various simple models like k-nearest neighbors or support vector machines can be trained to predict the latency of the represented plan~\cite{sparql_clustering_costmodel}.

\paragraph{Tree Convolution}
As the fixed-size algebraic and graph similarity vectors only encode the plan coarsely, the explicit query tree representation is a more powerful approach. For this, approaches like tree convolution can be used to process tree-structured data and, thus, capture all information in the representation~\cite{sparqltreeconvcost}. Such a model can also be trained in a supervised way using the tree representations of queries and the corresponding latency. In experiments, using a tree convolution with algebraic and graph-similarity features leads to a three-fold improvement over just using the coarse features.

\subsubsection{Limitations and Open Challenges}
Compared to approaches for relational databases, fewer approaches exist for learned   cost estimation over KGs. 
While existing approaches already display good results in terms of latency prediction accuracy, they can be further enhanced in terms of explicit and fine-grained representation of the query tree regarding operators and statistics about involved nodes and edges (e.g., using KG embeddings). 
The existing approaches also do not treat the problem of dynamic data or changes in the underlying hardware where the KG is stored, which can lead to significant changes in latency. 
For the latter, devising neuro-symbolic cardinality estimators is a promising direction, as they allow for decoupling the impact on the cost of the algebraic aspects of the query plan and the technical characteristics of the hardware used to store and manage the KG.

\begin{figure}[t]
    \centering
    \includegraphics[width=1\textwidth]{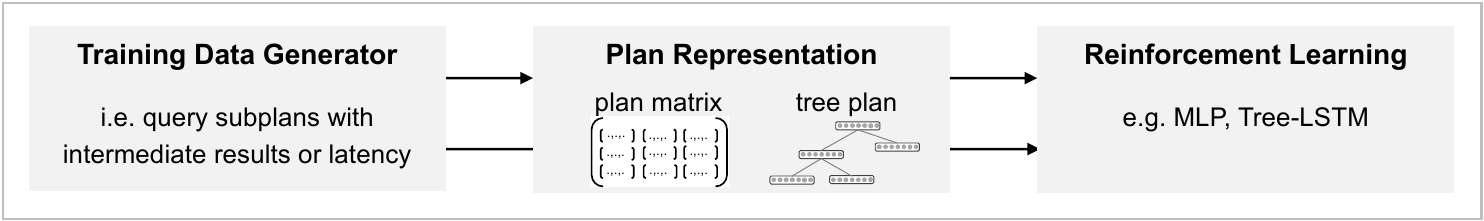}
    \caption{Architecture of learned plan traversal models for KG}
    \label{fig:kg_planner_architecture}
\end{figure}

\subsection{Learned Plan Traversal}
\label{sec:kg_learned_plan_traversal} 



Learned end-to-end models for plan traversal in KG aim to learn a greedy heuristic for operator ordering that minimizes the query execution latency. 
Similar to relational databases, existing works in KG also frame this task as a reinforcement learning (RL) problem. 
Figure~\ref{fig:kg_planner_architecture} shows the general architecture of a learned planner. 
In the RL models, the current \textit{state} of the environment is a partially completed plan, and the model can choose as \textit{actions} viable query operators that have not yet been included in the plan. 
The model is trained to optimize for a reward that correlates with the cost of the produced query plans, e.g., the number of intermediate results or the plan latency.

Unlike the relational case, the heterogeneity of KGs prevents encoding the complete database into the query plan representation, as done in the work by Marcus et al.~\cite{marcus2018deep}. Furthermore, queries in KGs usually contain a significantly larger number of operations (especially joins) than queries over relational databases, making them harder to optimize.

\subsubsection{Plan Representation}
Learned plan traversal models for KGs need to represent (sub)plans in a suitable way for reinforcement learning (RL) models. The following paragraphs discuss two common methods for encoding plans: matrix representation and query tree representation. 

\paragraph{Matrix Representation}
One way to encode a (sub)plan is using a square matrix, where each entry encodes an atomic operator~\cite{rejoop}, e.g., a triple pattern for SPARQL queries or a subgraph pattern for other KG data models. 
Entries on the diagonal represent operators still missing in the plan, while entries on the off-diagonal either represent performed operators or placeholders indicating that the operator can or cannot be performed. 
In the case of triples, they can be encoded using \textsc{id}s for the nodes, edges, and variables. 
While simple, this representation does not explicitly model the query tree and has no semantic information about the KG terms.  

\paragraph{Query Tree Representation}
Like learned cost models, another way is to explicitly represent the plan as a tree, where the nodes can be enhanced with nodes and edges that occur in the query, potential KG embeddings thereof~\cite{sparql_rtos}, and subplan cardinalities.

\subsubsection{Neural Architectures}
Learned planners leverage reinforcement learning to enhance and devise efficient query plans. 
The following paragraphs describe two specific neural network approaches -- Multilayer Perceptrons (MLPs) and Tree-LSTMs -- employed to predict the next operation in a query plan, each with unique capabilities and advantages.

\paragraph{Multilayer Perceptron}
One approach to predict the next join operation is to use an MLP based on the flattened matrix representation explained above~\cite{rejoop}. This simple model can then be trained using popular reinforcement learning approaches, such as Proximal Policy Optimization or Q-Learning, to predict the next operator to be included in the plan such that the overall number of results or the query latency is minimized. 

\paragraph{Tree-LSTM}
As in learned cost models, another approach is to use an architecture that can directly process a tree structure. Apart from the already introduced tree convolution is to use a Tree-LSTM that can similarly process an explicit tree structure to derive an output, in this case, the next join operation to be performed~\cite{sparql_rtos}. This model can be trained equally to the MLP but has a higher expressive power. 

\subsubsection{Limitations and Open Challenges}
While current approaches for learned plan traversal show results comparable to dynamic programming while attaining a linear runtime, issues similar to learned cardinality and cost methods remain. Reinforcement learning is typically sample inefficient, and thus, a large set of query plans and their corresponding runtime is needed. Furthermore, changes to the data or the database setup make resampling the dataset and retraining the models necessary. Future approaches thus need to enhance how such models are optimized, e.g., by using fast and efficient representation learning and incorporating fast-to-obtain statistics to enable faster and less frequent learning of the models themselves. In this regard, approaches not based on reinforcement learning are also a promising avenue.









\section{Challenges and Future Directions for Neuro-Symbolic Query Optimization}
\label{sec:future_directions} 

This chapter has presented the architecture of neuro-symbolic optimizers, where one or several symbolic components of the optimizer are entirely replaced by neural components. 
Another paradigm of using neural models during query optimization is to build a hybrid architecture where neural networks have a supporting role while the decision-making process is still delegated to the symbolic components.   
For instance, neural networks could be used to estimate the cost of a plan, which is then validated or adjusted using traditional rule-based or statistical methods. This hybrid approach can offer the benefits of neural models (e.g., learning from data) while maintaining the interpretability of symbolic methods. 
This hybrid architecture, however, introduces overhead during query optimization as it requires executing additional components, which hinders the overall query performance. 
For this reason, the goal is then to build neuro-symbolic optimizers where the neural components are first-class citizens. 
Besides the technical challenges already discussed in \S\ref{sec:ns-optimizer-kg}, in the following, we discuss further challenges and future directions to deploy neuro-symbolic query optimizers in real-world systems.  

\smallbreak
\paragraph{Generalization}
One of the key challenges in neuro-symbolic query optimization is ensuring that the models generalize well across different types of queries, dataset structures, and data distributions. 
A model that performs well on one type of workload may not necessarily generalize to others, particularly in the presence of novel query patterns or changes in the underlying dataset.

\smallbreak
\smallbreak
\smallbreak
\textit{Future Directions:}
\begin{itemize}
\item Hybrid data-query neural architectures: Data- and query-driven models for query optimization present inherent strengths and limitations. In particular, data-driven models can be challenging to scale up to large datasets, while query-driven models may overfit the training workload. Hence, hybrid data-query architectures can help the query-driven model generalize beyond the specific queries seen during training. 

\item Robust feature extraction: Improving the feature extraction process to ensure that the neural network can capture the most relevant and generalizable information of the query and database, which will reduce overfitting to specific types of queries. 

\item Meta-learning: Explore meta-learning approaches that enable the neural components in the query optimizer to learn how to learn. This enables quick adaption to new scenarios with minimal retraining.
\end{itemize}

\paragraph{Training Overhead and Scalability to Large Datasets}
Compared to traditional optimizer components that compute summaries or statistics, training neural components often requires a large amount of input data and significant computational resources. 
The training process can be time-consuming, especially for large datasets and complex queries. 
In particular, data-driven neural architectures may suffer from scalability over large datasets, as learning from KGs with billions of statements is currently impractical.   
Furthermore, especially in the case of query-driven neural architectures, training data (i.e., a query workload) that is representative and diverse is not always available. 

\smallbreak
\textit{Future Directions:}
\begin{itemize}
    \item Training data generators: Develop advanced data generators to efficiently produce large training data, e.g., KG with certain structures or queries and their costs, that is diverse and representative of various dataset states and query patterns. Furthermore, these data generators should be able to resemble realistic query workloads, incorporating common query patterns, edge cases, and noise that the optimizer might encounter in production environments.

    \item Hybrid data-query neural architectures: Hybrid architectures can incorporate query-level or other statistics during learning. This relieves the data-driven model from learning from large datasets and increases training efficiency.

    \item Efficient training algorithms: Research on more efficient training algorithms tailored to (knowledge graph) databases. For example, leveraging few-shot or transfer learning would reduce the overhead associated with training. 
    
    \item Incremental learning: Focus on developing incremental learning methods that allow the optimizer to adapt and update its knowledge without retraining the neural components whenever new data or queries are introduced.
\end{itemize}

\paragraph{Interpretability}
In traditional query optimizers, the (symbolic) components are inherently interpretable.  
This is an important feature for database administrators who need to understand the system's decision-making process to perform fine-tuning or correcting actions.  
Yet, neural components often function as opaque boxes. 
This lack of interpretability makes it difficult for users to understand, trust, and diagnose the optimizer's behavior, particularly when it makes unexpected or suboptimal decisions.

\smallbreak
\smallbreak
\smallbreak
\textit{Future Directions:}
\begin{itemize}
    \item Hybrid explainability techniques: Develop methods to provide explanations for the neural components' decision-making processes during query optimization. These methods may translate neural predictions into symbolic rules or use attention mechanisms that highlight important features of the queries or the dataset that impacted the component's output.

    \item User-centric tools: Build tools that allow users to query the decision process of neuro-symbolic components, thus improving trust and transparency. 
\end{itemize}

\paragraph{Uncertainty}
Traditional optimizers are sensitive to misestimations and other errors that introduce uncertainty along the query optimization process and lead to suboptimal plans.  
Besides these inaccuracies, neuro-symbolic optimizers can also be sensitive to noise, variations in the input, or biases introduced during learning, e.g., underrepresented query types in the training workload. This uncertainty can degrade the query performance, making it crucial to ensure that these systems are robust and can quantify their own uncertainty in decision-making.

\smallbreak
\textit{Future Directions:}
\begin{itemize}
\item Uncertainty quantification: Develop techniques to quantify and communicate the uncertainty in the neural predictions, which could be used to trigger fallbacks to symbolic, reliable methods when uncertainty is high.

\item Neuro-symbolic robust query optimization: Robust query optimization for relational databases~\cite{markl2004robust} and KGs~\cite{DBLP:conf/semweb/HelingA20,DBLP:journals/semweb/HelingA22} aims at identifying alternative, potentially non-optimal plans that are less sensitive or ``robust'' to optimization errors or unexpected adverse runtime conditions. Implementing neural-symbolic robust techniques requires incorporating adversarial examples during training to understand the vulnerabilities of the neural components and devise efficient plans.  

\item  Neuro-symbolic adaptive query processing: Adaptive query processing (AQP)~\cite{deshpande2007adaptive} goes beyond query optimization. AQP adjusts the behavior of the query execution schedulers when they are affected by unexpected conditions (i.e., uncertainty at runtime), e.g., suboptimal plans due to unexpected data correlations or environment unpredictability. For this, AQP implements monitoring mechanisms that collect performance feedback at runtime and take corrective actions, e.g., by changing the plan on the fly.  
Existing adaptive learning-based approaches~\cite{avnur2000eddies,acosta2015networks} implement simple techniques that do not scale well to different uncertainty types.   
Therefore, novel neuro-symbolic AQP architectures that can learn adaptivity policies are required to cope with runtime uncertainty during query processing.     
\end{itemize}

\section{Conclusion}
\label{sec:conclusion}
In this chapter, we explored neuro-symbolic architectures for query optimization in knowledge graphs.
As a relatively novel research area within database management and knowledge graph querying, this topic presents a unique intersection of symbolic reasoning and neural computation. 
As the literature shows, recent query optimization solutions have focused on replacing symbolic components, e.g., a cardinality estimator, with neural ones. Yet, this chapter highlighted the potential benefits of integrating symbolic and neural elements within a query optimizer. 
This interplay allows the optimizer to more effectively navigate the search space and devise efficient execution plans. 

However, despite these promising advancements, several significant challenges remain in deploying neuro-symbolic query optimizers in real-world systems. 
Among these challenges are the difficulties in generalizing the neural models to handle new query workloads and dynamic updates to the dataset, which are common requirements in real-world knowledge graphs. 
Furthermore, when applied to large datasets, the training overhead and scalability issues of machine learning models pose substantial obstacles. 
Lastly, incorporating neural components into query optimizers introduces issues of interpretability and uncertainty, which hamper the tasks related to database administration and must be managed to ensure reliable and predictable query optimization and execution.

\bibliographystyle{vancouver}
\bibliography{references}

\end{document}